\documentclass[aps,pra,twocolumn,showpacs,groupedaddress,nofootinbib]{revtex4-1}
\usepackage{bm,bbm,amsmath,bbold,amssymb,amsbsy,graphicx,subfigure,txfonts,color,hyperref,xfrac}
\usepackage[latin1]{inputenc}
\newcommand{\kebra}[2]{\vert{#1}\rangle\langle{#2}\vert}

\newcommand{\proj}[1]{\ket{#1}\bra{#1}}

\newcommand{\be}{\begin{equation}}
\newcommand{\ee}{\end{equation}}
\newcommand{\ben}{\begin{eqnarray}}
\newcommand{\een}{\end{eqnarray}}
\newcommand{\bes}{\begin{subequations}}
\newcommand{\ees}{\end{subequations}}

\newcommand{\tr}{\text{tr}}
\newcommand{\gr}[1]{\boldsymbol{#1}}
\newcommand{\ket}[1]{|#1\rangle}
\newcommand{\bra}[1]{\langle#1|}

\newcommand{\sig}{{\gr\sigma}}
\newcommand{\gam}{\boldsymbol{\gamma}}

\begin{document}
\title{Genuine multipartite nonlocality of permutationally invariant Gaussian states}

\date{December 19, 2016}
\author{Buqing Xu}
\email{ppxbx1@nottingham.ac.uk}
\affiliation{Centre for the Mathematics and Theoretical Physics of Quantum Non-Equilibrium Systems,
School of Mathematical Sciences, The University of Nottingham,
University Park, Nottingham NG7 2RD, United Kingdom}

\author{Tommaso Tufarelli}
\email{tommaso.tufarelli@nottingham.ac.uk}
\affiliation{Centre for the Mathematics and Theoretical Physics of Quantum Non-Equilibrium Systems,
School of Mathematical Sciences, The University of Nottingham,
University Park, Nottingham NG7 2RD, United Kingdom}


\author{Gerardo Adesso}
\email{gerardo.adesso@nottingham.ac.uk}
\affiliation{Centre for the Mathematics and Theoretical Physics of Quantum Non-Equilibrium Systems,
School of Mathematical Sciences, The University of Nottingham,
University Park, Nottingham NG7 2RD, United Kingdom}

\begin{abstract}
We investigate genuine multipartite nonlocality of pure permutationally invariant multimode Gaussian states of continuous variable systems, as detected by the violation of Svetlichny inequality. We identify the phase space settings leading to the largest violation of the inequality when using displaced parity measurements, distinguishing our results between the cases of even and odd total number of modes. We further consider pseudospin measurements and show that, for three-mode states with asymptotically large squeezing degree, particular settings of these measurements allow one to approach the maximum violation of Svetlichny inequality allowed by quantum mechanics. This indicates that the strongest manifestation of genuine multipartite quantum nonlocality is in principle verifiable on Gaussian states.
\end{abstract}

\pacs{03.65.Ud, 03.67.Mn, 42.50.Dv}

\maketitle

\section{Introduction}\label{sec:Intro}

Quantum mechanical systems can be correlated in ways stronger than classical ones. The characterization and exploitation of such correlations is enabling the development of a wealth of quantum technologies, set to revolutionize information and communication and other industrial sectors. Bell nonlocality is the strongest form of quantum correlations \cite{bellrev}. It manifests itself when two or more subsystems are in an entangled state \cite{entanglement} and additionally fulfill the more stringent condition that the outcomes of local measurements on each subsystem cannot be explained by using a local hidden variable model \cite{EPR35,bell}. This entails that the entanglement distributed among the subsystems, which may be located in different laboratories,  can be verified without any need for characterizing, or trusting, the measurement apparatuses available in each laboratory \cite{bellrev}. In turn, this ensures that the states exhibiting Bell nonlocality can be useful as resources for fully device-independent quantum communication, including in particular unconditionally secure quantum key distribution \cite{qkd}.

Nonlocal correlations can be detected by the violation of Bell-type inequalities \cite{bell}.   While a great deal of attention has been devoted to the study of Bell inequalities in bipartite systems \cite{EPR35,bell,chsh,werner89,bellrev}, including most recently the first loophole-free experimental demonstrations \cite{loopholefree1,loopholefree2,loopholefree3},  a few criteria have been formulated for the verification of Bell nonlocality in multipartite systems as well \cite{merminklyshko,svelt,roberts,bancalprl,sveltghzw,expsvelt,othermulti,operationonloc,bancal,bellrev}. However, the concept of {\em genuine} multipartite nonlocality has been formalized only recently from an operational point of view \cite{operationonloc,bancal} and its full exploration remains challenging both theoretically and experimentally.

In this paper we present a theoretical study of genuine multipartite nonlocality in multimode Gaussian states of infinite-dimensional systems. These states, which include squeezed and thermal states of quantized electromagnetic fields, are the theoretical pillars and the resources of choice for a number of applications in quantum information with continuous variables \cite{barnett,brareview,gaussreview,ourreview,book}. Their nonlocal properties have been explored in a few papers  \cite{binning,homobell,kb,jeong,bramermin,chen,ferraroparis,maurorefs,jie,mauro,jie2,samy}, albeit mostly limited to two or three modes.
 We investigate genuine multipartite nonlocality as revealed by the violation of an inequality first proposed for tripartite states by Svetlichny \cite{svelt}. Such an inequality can be violated e.g.~by both Greenberger-Horne-Zeilinger (GHZ) and $W$ classes of states for three qubits \cite{sveltghzw,expsvelt}, and it stands as the conventional witness of genuine nonlocality when all three parties perform two measurements with two outcomes each \cite{bancal}, even though a large number of weaker inequalities revealing genuine nonlocality in more general settings have been constructed more recently \cite{operationonloc,bancal}. While originally formulated for dichotomic observables, Bell-type inequalities such as the Svetlichny one can be tested for continuous variable systems either by binning outcomes of observables with continuous spectrum (typically Gaussianity-preserving measurements such as homodyne detection) \cite{binning,homobell}, or by considering directly operators with a discrete spectrum \cite{kb,chen}. The first kind of approach is of no use with Gaussian states, since their defining property of admitting a phase space description in terms of Gaussian (i.e., classical-like) Wigner distributions entails that all the results of homodyne detections can be fully explained by a local hidden variable model, even for entangled states. However, Gaussian states do exhibit trademark nonclassical and nonlocal features, as revealed by violations of Bell-type inequalities using observables of the second kind (which do not preserve Gaussianity), including most importantly displaced parity \cite{kb} and pseudospin \cite{chen}.

 Here we investigate the maximum violation of the Svetlichny inequality in pure permutationally invariant multimode Gaussian states, which can be seen as continuous variable analogs of multiqubit GHZ states \cite{network,vanlofuru,aoki03,ghzw,reidfuru}, by considering these two types of measurements. For displaced parity measurements, we extend the results of \cite{samy} from three to an arbitrary number of modes, providing a prescription to identify the phase space settings leading to the largest violations. However, we show that (as in the case of two and three modes) these violations do not reach the absolute maximum allowed by quantum mechanics, that is the multipartite analogue of the Tsirelson bound \cite{bellrev}. On the contrary, we provide substantial evidence that by using pseudospin operators one can approach such a maximum violation asymptotically in pure permutationally invariant three-mode Gaussian states. This result, which mirrors the case of bipartite nonlocality in two-mode squeezed Gaussian states \cite{chen}, demonstrates theoretically that maximum genuine multipartite quantum nonlocality is in fact attainable in continuous variable Gaussian states, provided arbitrarily large squeezing is available. Extensions of the latter analysis to an arbitrary number of modes are certainly possible but cumbersome, since our treatment of pseudospin measurements relies on the explicit expansion of Gaussian states in the Fock basis. Such generalizations are thus left for future work.

This paper is organized as follows. Section~\ref{sec:Gauss} recalls essential concepts and tools for continuous variable Gaussian states and their mathematical description. Section~\ref{sec:Svetl} introduces basic notions in Bell nonlocality and the Svetlichny inequality criterion for genuine multipartite nonlocality. Section~\ref{sec:Parity} presents our results on the optimal violation of Svetlichny inequality for multimode permutationally invariant Gaussian states using displaced parity measurements. Section~\ref{sec:Pseudo} contains our study of genuine tripartite nonlocality, up to the maximum quantum bound, for three-mode permutationally invariant Gaussian states using pseudospin measurements. Section~\ref{sec:Concl} concludes the manuscript with a brief outlook.


\section{Gaussian preliminaries}\label{sec:Gauss}
We consider a $n$-mode continuous variable system, characterized by an infinite-dimensional Hilbert space resulting from the tensor product of the Fock spaces of each mode. This describes, for instance, a collection of $n$ quantum harmonic oscillators.  The quadrature operators for each mode are collected in the vector $\hat{\boldsymbol{R}} = (\hat q^1, \hat p^1, \hat q^2, \hat p^2, \ldots, \hat q^n, \hat p^n)^{\sf T}$, so that the canonical commutation relations can be written compactly as $[\hat R_j, \hat R_k] = i \left(\boldsymbol\omega^{\oplus n}\right)_{j,k}$ with $\boldsymbol\omega = {{\ 0 \ \ 1} \choose {-1 \ 0}}$ being the symplectic form.
A {\it Gaussian state} $\rho$ is represented by a Gaussian   phase space Wigner distribution \cite{ourreview},
\begin{equation}
\label{wigner}
W_\rho(\boldsymbol{\xi}) = \frac{1}{\pi^n \sqrt{\det{\sig}}} \exp\big[-(\boldsymbol{\xi}-\boldsymbol{\delta})^{\sf T} \sig^{-1} (\boldsymbol{\xi}-\boldsymbol{\delta})\big]\,,
\end{equation}
where $\boldsymbol{\xi} \in \mathbb{R}^{2n}$ is a phase space coordinate vector, $\boldsymbol{\delta} = \langle \hat{\boldsymbol{R}}\rangle$ is the vector of first moments of the canonical variables, and $\sig$ is the covariance matrix  of second moments, whose entries are  $\sigma_{j,k}=\text{tr}[\rho \{\hat{R}_j-\delta_j,\hat{R}_k-\delta_k\}_+]$, with  $\{\cdot,\cdot\}_+$ denoting the anticommutator. The first moments can be adjusted by local displacements, which have no effect on the correlations among the modes, therefore from now on we will assume $\boldsymbol{\delta} = \boldsymbol 0$ without any loss of generality. We will then focus on the covariance matrix $\sig$, which encodes all the correlation properties of a Gaussian state $\rho$. Any real, symmetric, $2n \times 2n$ matrix $\sig$ needs to satisfy the condition
\begin{equation}\label{bona}
\sig + i \boldsymbol\omega^{\oplus n} \geq 0\,,
\end{equation}
in order to be a valid covariance matrix for a physical state $\rho$ \cite{simon94}. In passing, note how Eq.~\eqref{bona} may be seen as a generalization of the Robertson-Schr\"odinger uncertainty principle. The purity of a Gaussian state $\rho$  is given simply by $\mu(\rho) = \tr\,\rho^2 = (\det\sig)^{-\frac12}$, so that a pure Gaussian state  $\rho = \ket{\psi}\!\bra{\psi}$ has a covariance matrix $\sig$ with $\det\sig=1$, saturating the above matrix inequality (\ref{bona}).

We will focus our analysis on pure permutationally invariant $n$-mode Gaussian states \cite{ghzw}, which are known as the continuous variable counterparts of both GHZ and $W$ states of $n$ qubits, as they maximize both the genuine $n$-partite entanglement and the residual bipartite entanglement between any pair of modes, within the set of Gaussian states \cite{3mpra,contangle}. These states, often referred to as continuous variable GHZ-like states \cite{vanlofuru,aoki03,reidfuru}, have been investigated theoretically and experimentally as useful resources for multipartite teleportation networks \cite{network,telepoppy,naturusawa}, error correcting codes \cite{aoki}, and cryptographic protocols such as quantum secret sharing \cite{sharing,sharingyu,armstrong} and Byzantine agreement \cite{byz}. In the following we will show that they are very good candidates to reveal strong manifestations of genuine multipartite nonlocality by means of suitable measurements.

Up to local unitaries, the covariance matrix of these Gaussian states can be written in the following normal form in terms of $2 \times 2$ subblocks \cite{ghzw},
\begin{equation}
\label{sigma}
\sig=\left(
\begin{array}{ccccc}
\boldsymbol{\alpha} & \gam   & \gam & \cdots&  \gam  \\
\gam & \boldsymbol\alpha &\gam  & \cdots & \gam \\
 \gam   & \gam  & \ddots & \cdots & \gam \\
 \vdots & \vdots & \vdots &\ddots  & \gam  \\
\gam & \gam & \gam & \gam & \boldsymbol\alpha
 \end{array}
\right)\,
\end{equation}
where $\boldsymbol{\alpha} = {\rm diag}(a,\,a)$ and $\gam = {\rm diag}(z^+_n,\, z^-_n)$, with $a \geq 1$ and
\begin{equation}\label{zn}
z^\pm_n = \frac{(a^2-1)(n-2) \pm \sqrt{(a^2-1)(a^2 n^2 - (n-2)^2)}}{2a(n-1)}\,.
\end{equation}
These states are therefore entirely specified (up to local unitaries) by a single parameter, the local mixedness factor $a$, which can be accordingly expressed in terms of a single-mode squeezing degree $r$ needed to prepare the state via a network of beam splitters \cite{network,telepoppy,ghzw,contangle}.

\section{Svetlichny inequality}\label{sec:Svetl}

In the Bell scenario, nonlocality can be detected in the state of a composite system by allowing every party to perform a selection of different measurements on their subsystems, each with two or more possible outcomes. From the expectation values of the measured observables one then infers a correlation parameter, whose value is bounded if a  local hidden variable theory is assumed. Correlations exceeding the bound reveal the failure of local realism, that is, the presence of nonlocality. We refer the reader to \cite{bell,chsh,operationonloc,bancal,bellrev} and references therein for an updated account on the subject.

In order to introduce the Svetlichny inequality for genuine multipartite nonlocality \cite{svelt}, it is convenient to start with the bipartite case, i.e.~with the traditional Clauser-Horne-Shimony-Holt (CHSH) inequality \cite{chsh}. Suppose two experimenters Alice and Bob can each perform either one of two possible dichotomic measurements on their subsystem of a bipartite system. Say, Alice can measure her subsystem in either setting $A_0$ or $A_1$, with respective outcome $a_x$ ($x\in \{0,1\}$) and Bob can measure his subsystem in either setting $B_0$ or $B_1$ with respective outcome $b_y$ ($y \in \{0,1\}$). Here $a_0,a_1,b_0,b_1$ can take values $\pm 1$. Defining now $\langle a_x b_y \rangle = \sum_{a,b=\pm 1} a b\, {\mathbb P}(ab|xy)$ as the expectation value of the product of outcomes $ab$ for given measurement choices $x,y$, the Bell-CHSH parameter $M_2$ can then be written as follows \cite{chsh,bellrev}, adopting a convenient normalization \cite{roberts},
\begin{equation}\label{chsh}
M_2=\frac12 \left(\langle a_0 b_0 \rangle + \langle a_0 b_1 \rangle + \langle a_1 b_0 \rangle - \langle a_1 b_1 \rangle\right)\,.
\end{equation}
Assuming a model with a local hidden variable $\lambda$, according to which the expectation values can be factorized as $\langle a_x b_y \rangle = \int {\rm d}\lambda q(\lambda) \sum_a a \, {\mathbb P}(a|x,\lambda) \sum_b b \, {\mathbb P}(b|y,\lambda)$, it is straightforward to see that
\begin{equation}\label{chshineq}
M_2 \leq 1\,,
\end{equation}
which is known as the CHSH inequality.
However, if Alice and Bob share an entangled quantum state $\rho$, there exist measurement settings such that the parameter $M_2$ constructed from the expectation values of their experimental data violates the inequality (\ref{chshineq}), up to the Tsirelson bound
\begin{equation}\label{chshcirel}
M_2 \leq \sqrt 2\,,
\end{equation}
which represents the maximum violation compatible with quantum mechanics.

Consider now a tripartite system, and three observers Alice, Bob, and Charlie. The generalization of the CHSH inequality to this scenario is known as Mermin-Klyshko inequality \cite{merminklyshko}. Defining  $\langle a_x b_y c_z \rangle = \sum_{a,b,c=\pm 1} a b c \, {\mathbb P}(abc|xyz)$ as the expectation value of the product of outcomes $abc$ for given measurement choices $x,y,z$ of the three parties, the  Mermin-Klyshko parameter can be written as
\begin{equation}\label{mermin}
M_3= \frac12\left(\langle a_1 b_0 c_0 \rangle + \langle a_0 b_1 c_0 \rangle + \langle a_0 b_0 c_1 \rangle - \langle a_1 b_1 c_1 \rangle\right) \,.
\end{equation}
Once more, $M_3 \leq 1$ for any local hidden variable model, while $M_3$ can reach up to $\sqrt 2$ with entangled quantum states. However, a violation of the Mermin-Klyshko inequality can be achieved already by using only bipartite entangled states between any two of the three parties. To remedy this problem, one can define the  Svetlichny parameter $S_3$ for a tripartite system as $S_3 = (M_3 + \bar{M}_3)/2$, where $\bar{M}_3$ is obtained from $M_3$ by swapping the $0$'s and $1$'s in the settings $x,y,z$. Explicitly,
\begin{eqnarray}\label{svelt}
S_3&=&\frac14 \left(\langle a_1 b_0 c_0 \rangle + \langle a_0 b_1 c_0 \rangle + \langle a_0 b_0 c_1 \rangle - \langle a_1 b_1 c_1 \rangle \right. \nonumber \\
&& +   \left. \langle a_0 b_1 c_1 \rangle + \langle a_1 b_0 c_1 \rangle + \langle a_1 b_1 c_0 \rangle - \langle a_0 b_0 c_0 \rangle \right)\,.
\end{eqnarray}
In this way, a violation of the Svetlichny inequality
\begin{equation}\label{sveltineq}
S_3 \leq 1
\end{equation}
ensures that the correlations detected by Alice, Bob, and Charlie cannot be reproduced by any local hidden variable assigned
to the joint measurement of any two out of three parties. In this sense, a violation of (\ref{sveltineq}) reveals genuine tripartite nonlocality. Such a violation is possible using quantum mechanical states with genuine tripartite entanglement (i.e.~fully inseparable states), such as GHZ states of three qubits \cite{sveltghzw,expsvelt}, up to
\begin{equation}\label{sveltcirel}
S_3 \leq  \sqrt 2 \equiv S_3^Q\,,
\end{equation}
which defines the maximum allowed quantum violation $S_3^Q$.

The Svetlichny inequality can be generalized to detect genuine $n$-partite nonlocality. Consider a composite system partitioned into $n$ subsystems, each measured by an experimenter (labelled by the superscript $j=1,\ldots,n$) in two possible settings $O^j_{x_j}$ with respective outcomes $o^j_{x_j}$, where $x_j \in \{0,1\}$ and $o^j_{x_j} \in \{-1,1\}$. For $1\le k\le n-1$ the Mermin-Klyshko parameter can be defined recursively in a compact way  \cite{roberts},
\begin{equation}\label{merminn}
M_n = \frac12 M_{n-k}(M_k + \bar{M}_k) + \frac12 \bar{M}_{n-k}(M_k-\bar{M}_k)\,,
\end{equation}
where $M_1 = \langle o^1_0\rangle$, while  $M_2$ and $M_3$ are given by Eqs.~(\ref{chsh}) and (\ref{mermin}), respectively.
The Svetlichny parameter for arbitrary $n$ can be defined accordingly \cite{roberts},
\begin{equation}\label{sveltn}
S_n = \left\{
        \begin{array}{ll}
          M_n, & \hbox{even $n$;} \\
          \frac12(M_n + \bar{M}_n), & \hbox{odd $n$.}
        \end{array}
      \right.
\end{equation}
With the adopted normalization, violation of the generalized Svetlichny inequality
\begin{equation}\label{sveltineqn}
S_n \leq 1
\end{equation}
signals genuine $n$-partite nonlocality. Quantum mechanical violations are possible up to the maximum value
\begin{equation}\label{sveltcireln}
S_n \leq   2^{\frac12(n-1-({n\!\! \mod 2}))} \equiv S_n^Q\,,
\end{equation}
which generalizes the case $n=3$ reported in Eq.~(\ref{sveltcirel}).

In the following, we will focus on quantum states invariant under arbitrary permutations of the $n$ subsystems. In this case, let us denote by $E^m_n$ the expectation value of a product of joint measurements with $m$ settings  $x_j=1$ and $(n-m)$ settings  $x_j=0$; for example, $E^2_5$ can indicate a term like $\langle o^1_1 o^2_1 o^3_0 o^4_0 o^5_0 \rangle$, or any of its permutations. The $n$-partite Svetlichny parameter $S_n$ of Eq.~(\ref{sveltn}) acquires then the simple form
\begin{equation}\label{sveltnsym}
S_n = \frac{1}{2^{\left\lceil \frac{n}{2}\right\rceil }} \sum_{m=0}^n B^m_n E^m_n\,,
\end{equation}
where $B^m_n =  (-1)^{\left\lceil \frac{n-2 (m+1)}{4} \right\rceil } \binom{n}{m}$, with the binomial coefficient $\binom{n}{m} = \frac{n!}{m!(n-m)!}$, and  $\lceil \cdot \rceil$ denoting the ceiling function.

\section{Multipartite nonlocality with displaced parity measurements}\label{sec:Parity}

To test multipartite nonlocality in $n$-mode continuous variable systems, we first choose {\em displaced parity measurements} as the operators to be measured on each mode $j$ \cite{kb,jeong,mauro}. In the case of optical fields, the displaced parity observable $\hat{P}^j$ on mode $j$ can be measured by photon counting, preceded by a phase space displacement, the latter implemented e.g.~by beamsplitting the input mode with a tunable coherent field \cite{kb,expdispar}.
In formula,
\begin{equation}\label{dispari}
\hat{P}^j(\boldsymbol{\xi}^j_{x_j}) = \sum_{n=0}^{\infty} (-1)^n \ket{\boldsymbol{\xi}^j_{x_j},n}\bra{\boldsymbol{\xi}^j_{x_j},n}\,,
 \end{equation}
where $\ket{\boldsymbol{\xi}^j_{x_j},n}$ is the $n^{\rm th}$ Fock state of mode $j$, displaced by a phase space vector $\boldsymbol{\xi}^j_{x_j} \equiv (q^j_{x_j}, p^j_{x_j})$; notice that we are keeping a binary tag $x_j \in \{0,1\}$ to allow for the choice of two different phase space settings on each mode $j$.
These measurements have been implemented in recent Bell-type experiments with optical vortex beams \cite{expdispar2}.

It has been proven in \cite{kb} that, for an arbitrary (single-mode) quantum state $\rho_j$, the expectation value of a displaced parity operator $\hat{P}^j(\boldsymbol{\xi}^j_{x_j})$ is proportional to the Wigner distribution $W_\rho$ of $\rho$ evaluated in the phase space point with coordinates given by the setting $\boldsymbol{\xi}^j_{x_j}$, that is,    $\langle \hat{P}^j(\boldsymbol{\xi}^j_{x_j}) \rangle_{\rho_j} = \pi W_{\rho_j}(\boldsymbol{\xi}^j_{x_j})$. This result extends immediately to multimode states. We can then rewrite all the expectation values appearing in the Svetlichny parameter $S_n$, defined by Eq.~(\ref{sveltn}), in terms of the Wigner distribution of a $n$-mode quantum state $\rho$ evaluated at suitable phase space points. For instance, in the tripartite case, the first correlation function in Eq.~(\ref{svelt}) would read $\langle a_1 b_0 c_0 \rangle =  \pi^3 W_\rho(\boldsymbol{\xi}^1_1 \oplus \boldsymbol{\xi}^2_0 \oplus \boldsymbol{\xi}^3_0)$, and so on \cite{mauro,samy}. If $\rho$ is a multimode Gaussian state with zero first moments and covariance matrix $\sig$, its Wigner distribution is given by Eq.~(\ref{wigner}), and the Svetlichny parameter $S_n$ defined in Eq.~(\ref{sveltn}), for displaced parity measurements, depends only on the entries of the covariance matrix $\sig$, as well as on the measurement settings $\{\boldsymbol{\xi}^j_{x_j}\}_{x_j=0,1}^{j=1,\ldots, n}$.

In the following, we investigate the maximum value that the Svetlichny parameter $S_n$ can reach when performing local displaced parity measurements on $n$-mode permutationally invariant Gaussian states, whose covariance matrix is given by Eq.~(\ref{sigma}), and characterize the phase space settings leading to a violation of the Svetlichny inequality (\ref{sveltineqn}), that is, to a detection of genuine $n$-partite nonlocality.

Given the permutational symmetry of the states, we can assume that the binary set of available measurement settings is the same for each mode $j$, so that overall there will be  $m$ modes displaced by  $\boldsymbol{\xi}_1=(p_1,q_1)$, and $(n-m)$ modes displaced by $\boldsymbol{\xi}_0 = (p_0,q_0)$.  The expectation value $E^m_n$ of such a product of local displaced parity measurements can be then computed exactly, and takes the following expression
\begin{eqnarray}\label{Enm}
&&E^m_n = \exp \left[-z^-_n (q_0 (n-m)+m q_1)^2-z^+_n (p_0 (n-m)+m
   p_1)^2\right. \nonumber \\
&&\ \ +\left.(z^-_n -a) \left(q_0^2 (n-m)+m q_1^2\right)+(z^+_n -a)
\left(p_0^2 (n-m)+m p_1^2\right) \right]. \nonumber \\
\end{eqnarray}
Plugging the above into Eq.~(\ref{sveltnsym}), we get a compact formula for the Svetlichny parameter $S_n \equiv S_n(a; q_0,q_1,p_0,p_1)$.

\begin{figure*}[t!]
\subfigure[]{\includegraphics[width=4.2cm]{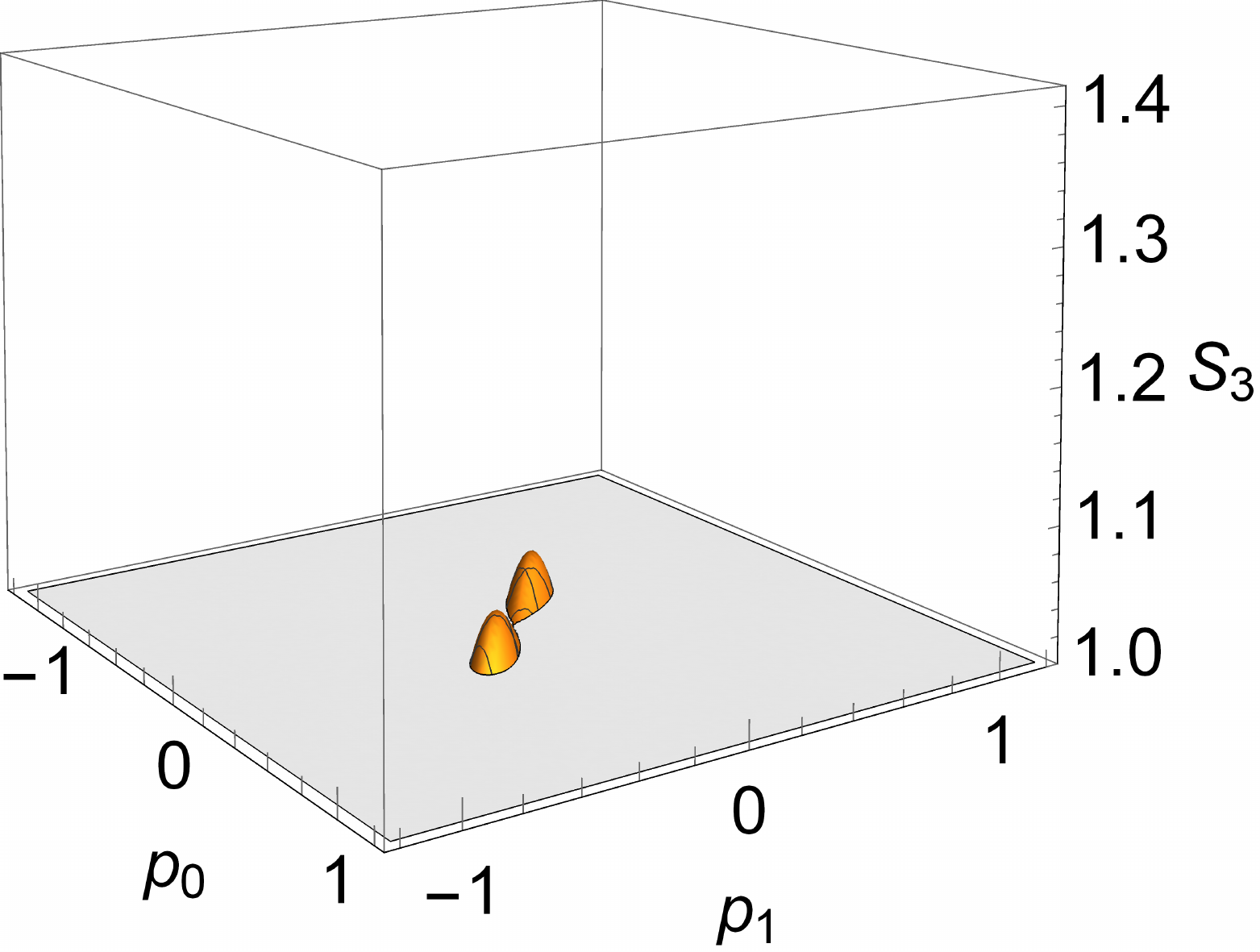}} \hspace*{.2cm}
\subfigure[]{\includegraphics[width=4.2cm]{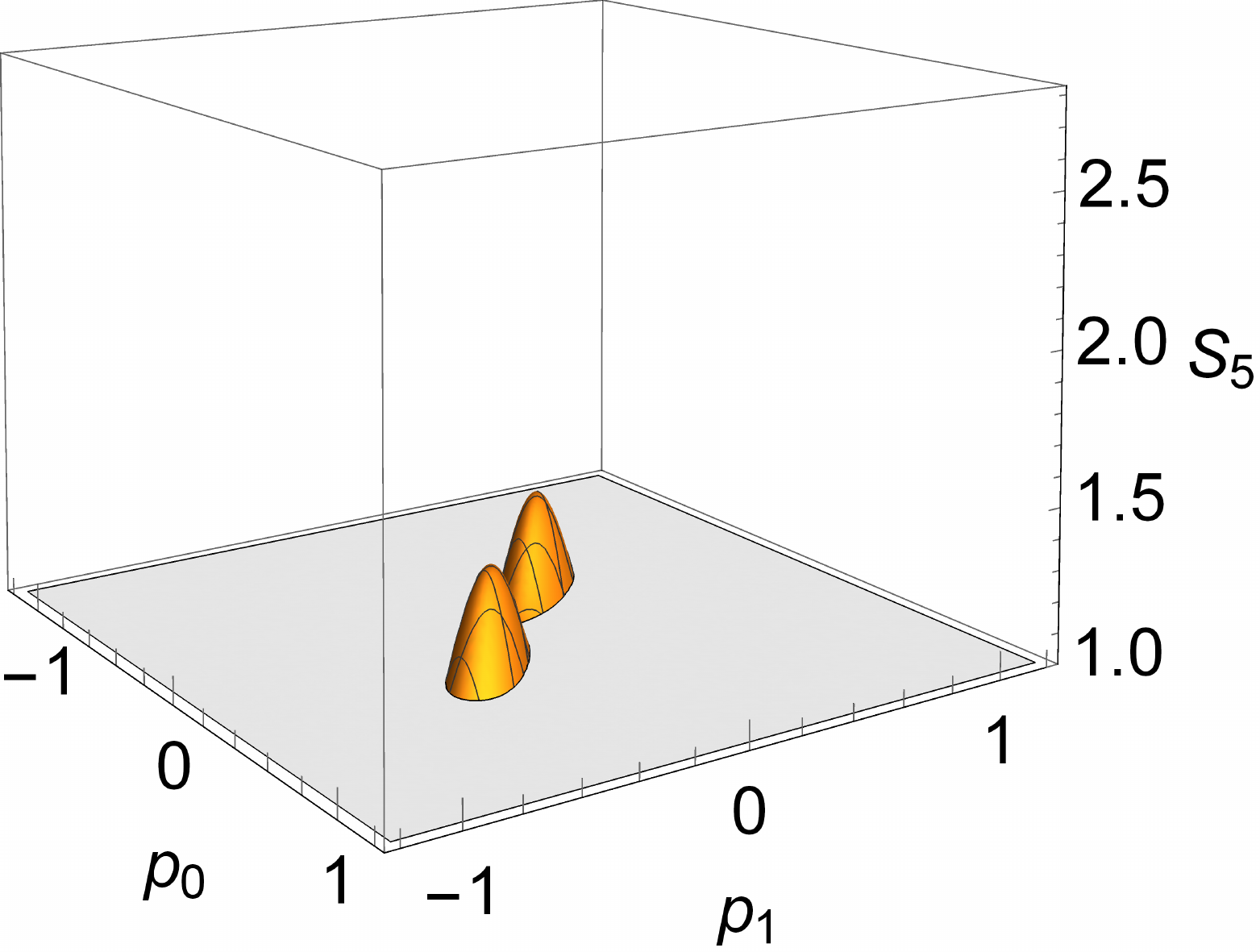}} \hspace*{.2cm}
\subfigure[]{\includegraphics[width=4.2cm]{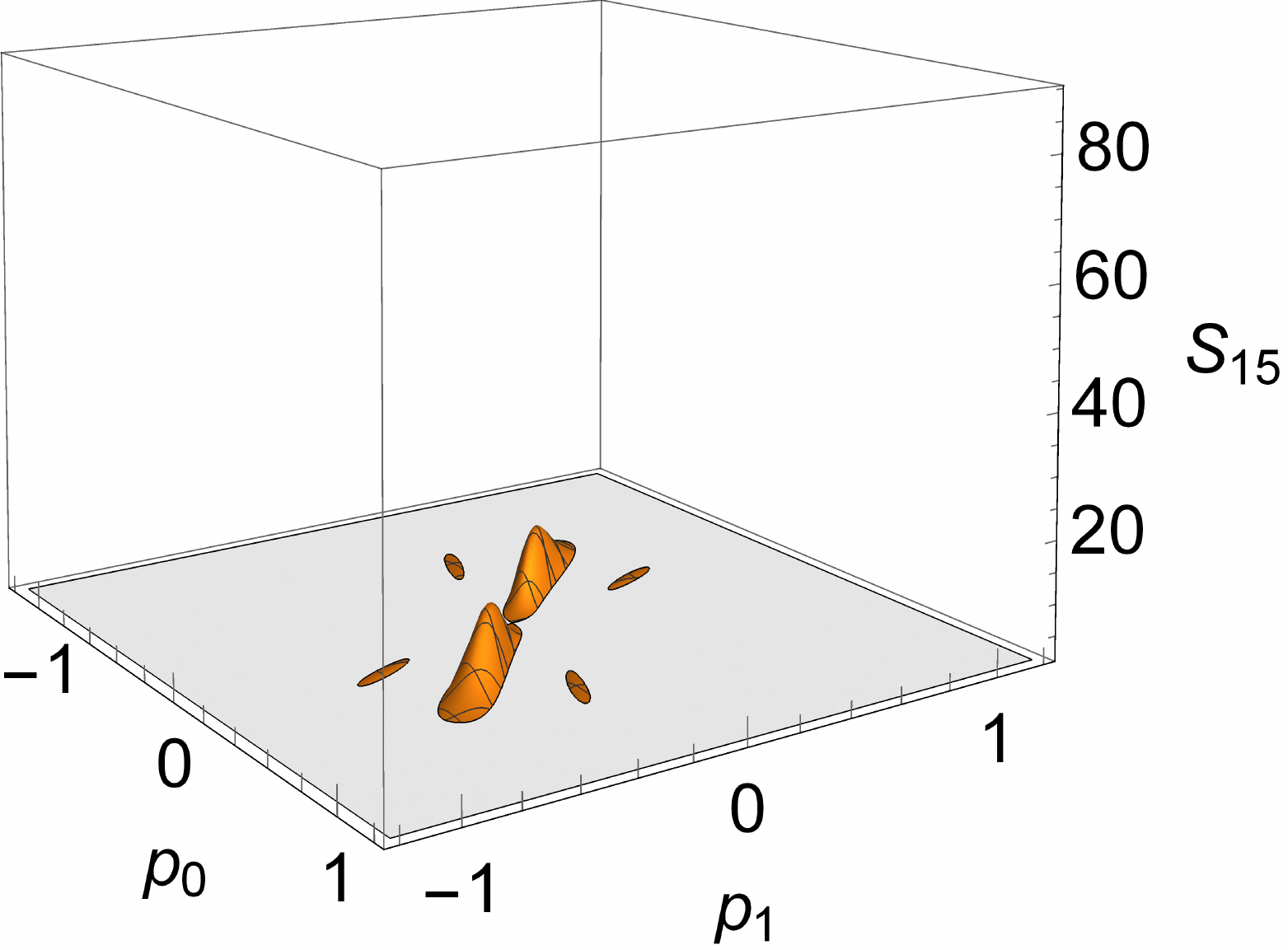}} \hspace*{.2cm}
\subfigure[]{\includegraphics[width=4.2cm]{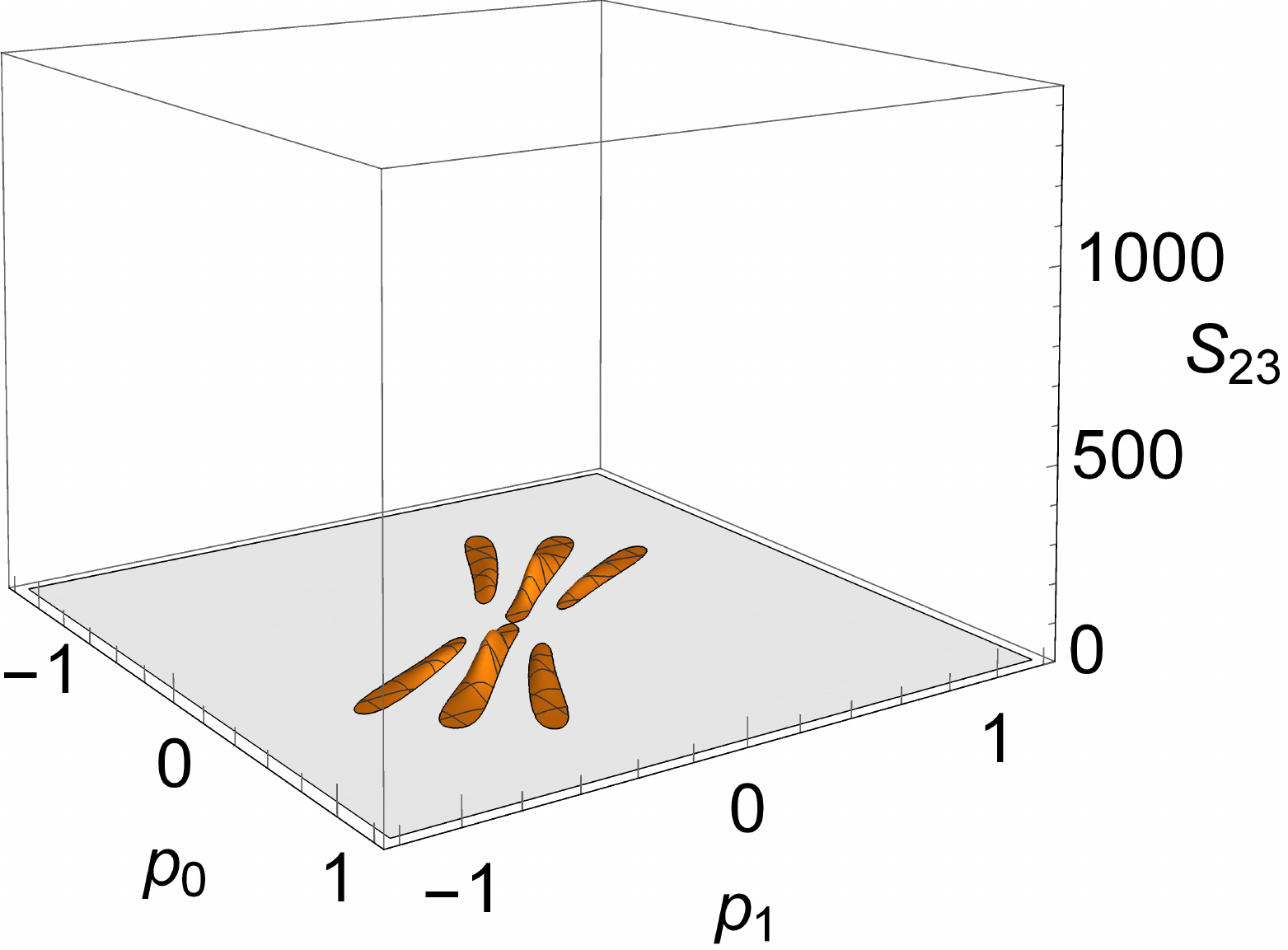}} \\[.2cm]
\subfigure[]{\includegraphics[width=4.2cm]{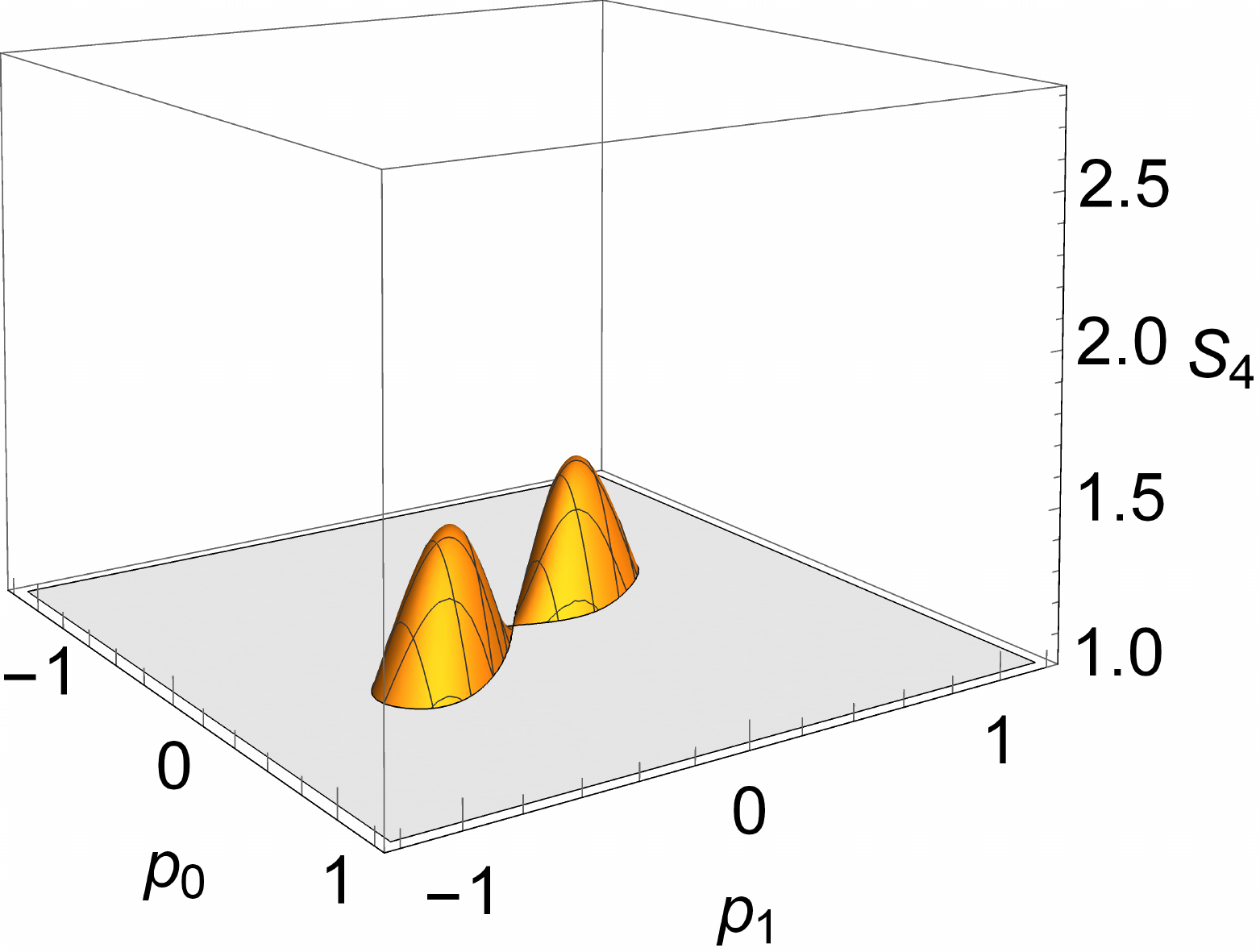}} \hspace*{.2cm}
\subfigure[]{\includegraphics[width=4.2cm]{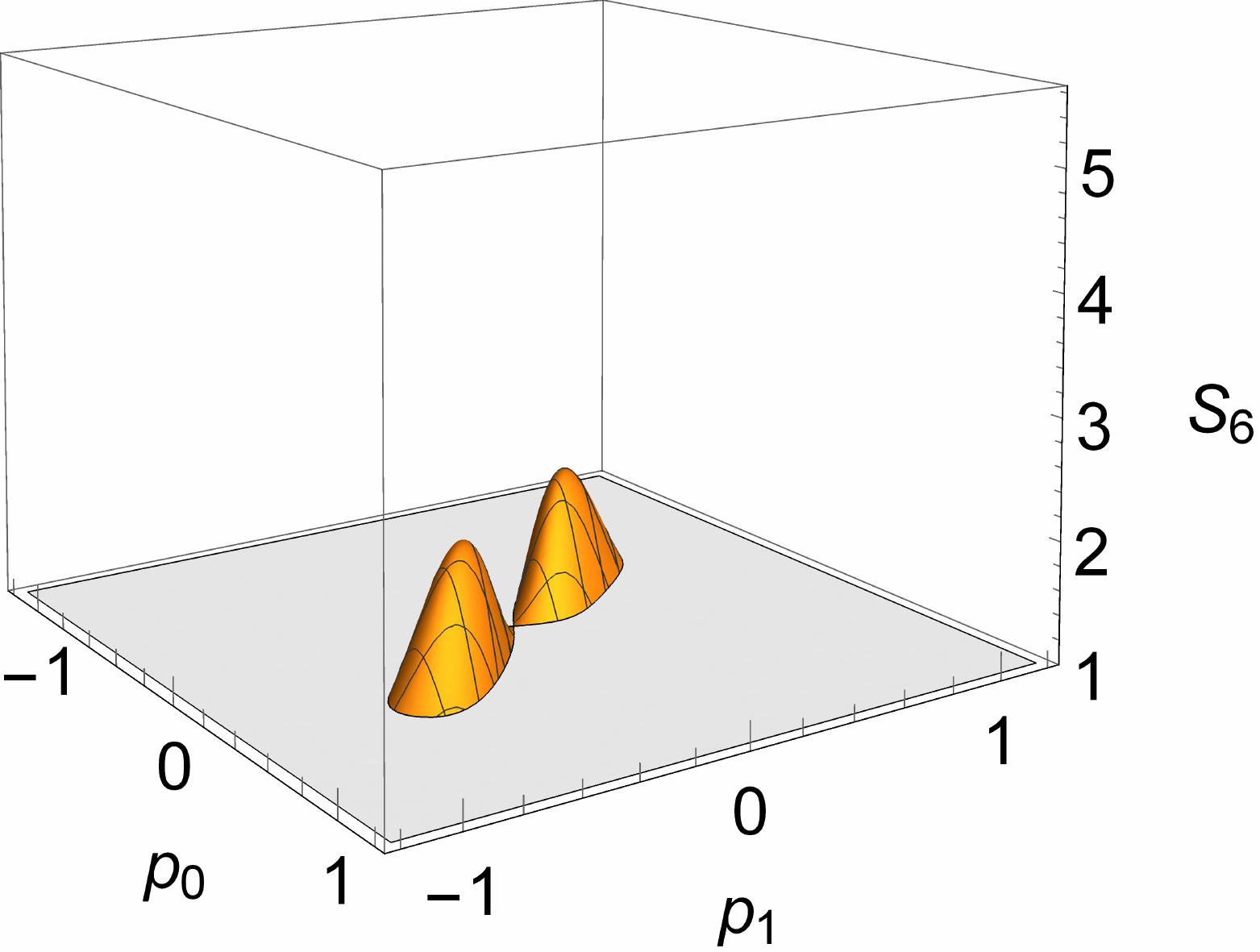}} \hspace*{.2cm}
\subfigure[]{\includegraphics[width=4.2cm]{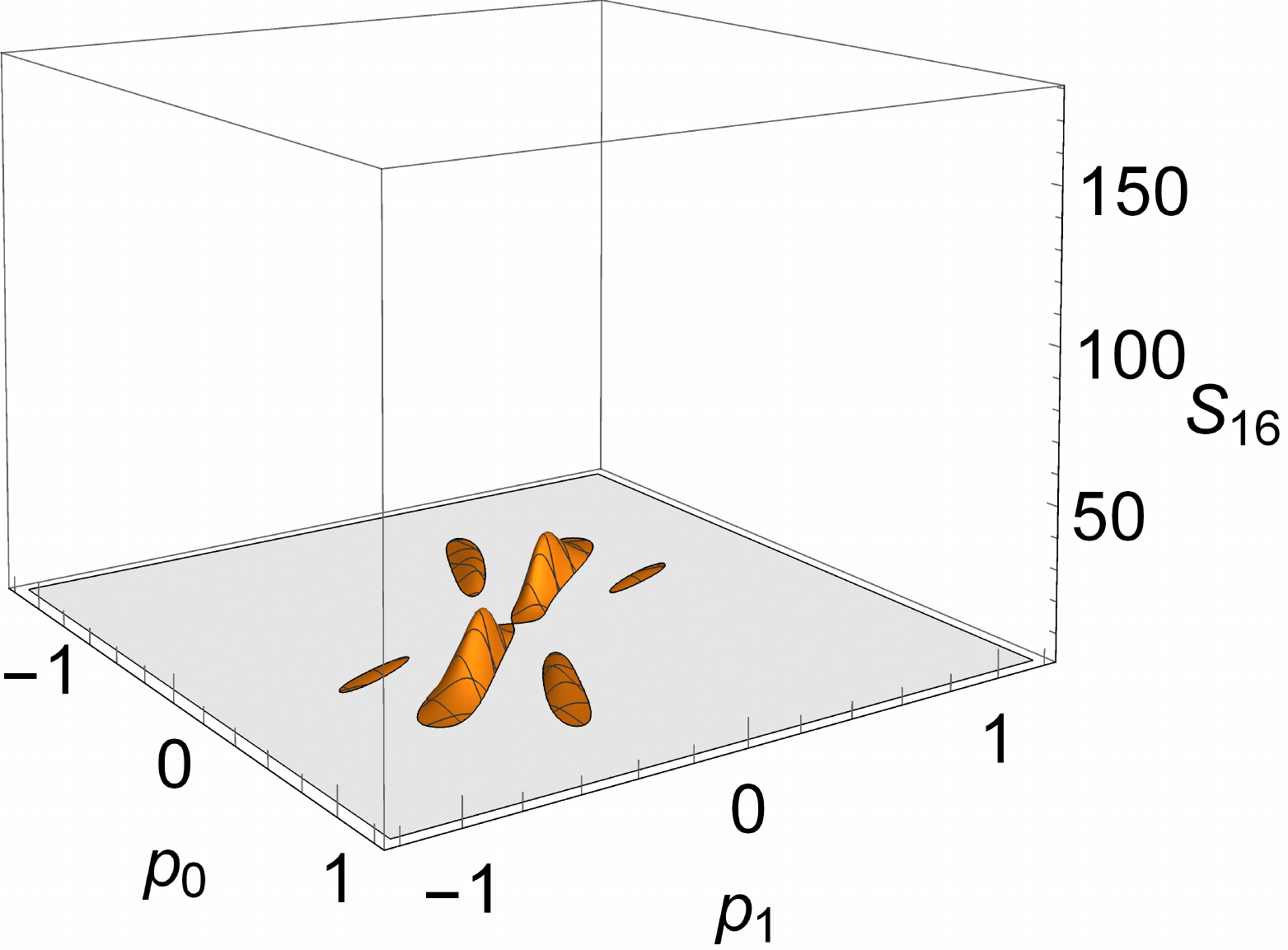}} \hspace*{.2cm}
\subfigure[]{\includegraphics[width=4.2cm]{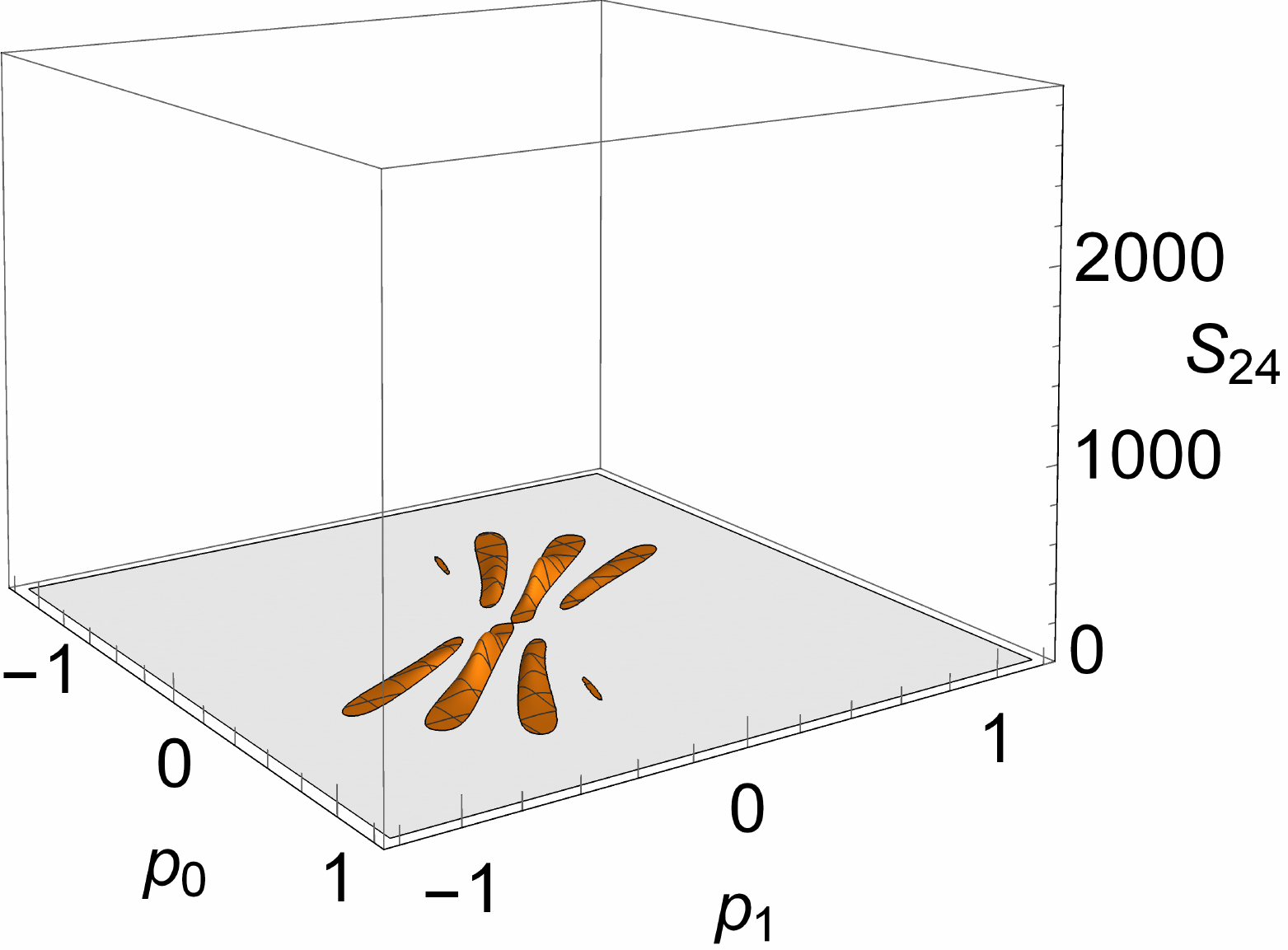}} \\[.2cm]
\caption{(Color online) Plots of the Svetlichny parameter $S_n(a)$ for pure permutationally invariant $n$-mode Gaussian states using  displaced parity measurements with $q_0=0=q_1$ and variable settings $p_0, p_1$, for $a=1.5$ and representative choices of  $n$. Top row: (a) $n=3$, (b) $n=5$, (c) $n=15$, and (d) $n=23$. Bottom row: (e) $n=4$, (f) $n=6$, (g) $n=16$, and (h) $n=24$. The vertical axis in each panel ranges from $1$ to the maximum quantum bound $S_n^Q$, so that only values of $S_n$ violating the Svetlichny inequality (\ref{sveltineqn}) are shown. All the plotted quantities are dimensionless.
\label{figeva}
}
\end{figure*}

Our next task is to optimize $S_n$ over the local phase space settings at given $a,n$, i.e., to find
\begin{equation}\label{sopt}
S^{\text{opt}}_n (a) = \max_{\{q_{0,1}, p_{0,1}\}} S_n(a; q_0,q_1,p_0,p_1)\,.
\end{equation}
By evaluating partial derivatives with respect to  $q_0$ and $q_1$, we can see that the setting $q_0=0=q_1$ yields a stationary point of $S_n$ for any $n$. A numerical analysis up to $n=30$ modes confirms that this choice of quadratures maximizes the Svetlichny parameter $S_n$. We are thus left to identify the optimal settings for $p_0$ and $p_1$, which are obtained by solving the following system of two transcendental equations,
\begin{eqnarray}\label{buqings}
0 & = & \sum_{m=0}^n [a m p_1 + z^+_n (m(n-m)p_0+m(m-1)p_1)]\nonumber \\
&&  \times B^m_n e^{-a(mp_1^2+(n-m)p_0^2)-z^+_n[2m(n-m)p_0p_1+m(m-1)p_1^2+(n-m)(n-m-1)p_0^2]} \,, \nonumber \\
&&  \\
0 & = & \sum_{m=0}^n [a (n-m) p_0 + z^+_n (m(n-m)p_1+(n-m)(n-m-1)p_0)] \nonumber \\
&& \times B^m_n e^{-a(mp_1^2+(n-m)p_0^2)-z^+_n[2m(n-m)p_0p_1+m(m-1)p_1^2+(n-m)(n-m-1)p_0^2]}\,. \nonumber
\end{eqnarray}
While an exact solution of these equations appears unfeasible for arbitrary $n$, we can make some general observations, supported by numerical analysis.

\begin{figure*}[tbh]
\subfigure[]{\includegraphics[width=8cm]{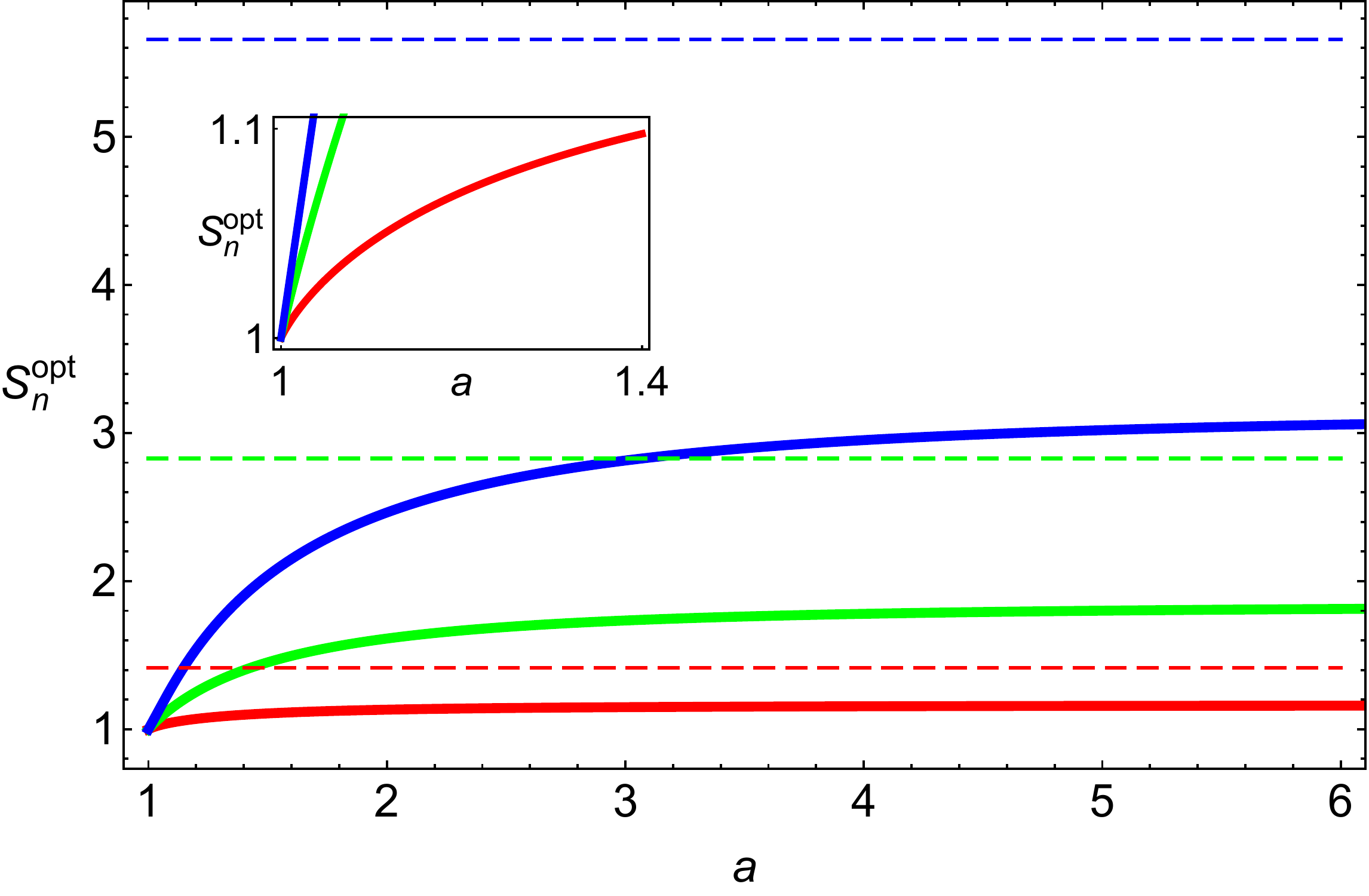}} \hspace*{1cm}
\subfigure[]{\includegraphics[width=8cm]{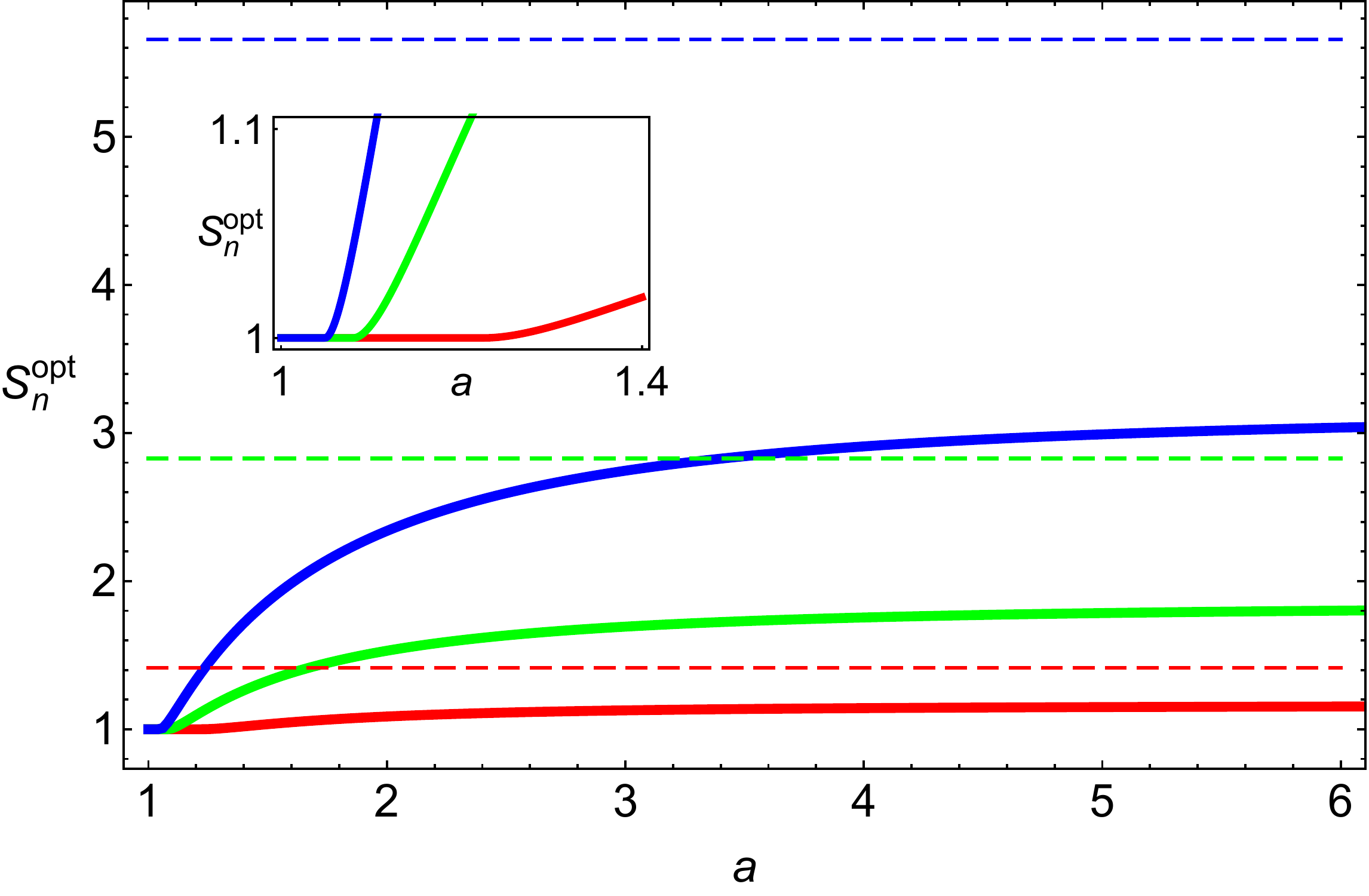}}
\caption{(Color online) Optimal Svetlichny parameter $S_n^{\text{opt}}$ (solid curves) for pure permutationally invariant $n$-mode Gaussian states using  displaced parity measurements, plotted  versus the covariance parameter $a$ for (from bottom to top): (a) $n=2$ (red online), $n=4$ (green online), $n=6$ (blue online), and (b) $n=3$ (red online), $n=5$ (green online), $n=7$ (blue online). The insets detail the regime of small $a$, showing that a threshold  for violations of the Svetlichny inequality exists in the odd $n$ case (b), but not in the even $n$ case (a).  In both panels, the dashed horizontal lines indicate the maximum value $S_n^Q$ of the Svetlichny parameter allowed by quantum mechanics, given by (from bottom to top) $S_2^Q=S_3^Q=\sqrt{2}$ (dashed red online), $S_4^Q=S_5^Q=2\sqrt{2}$ (dashed green online), and $S_6^Q=S_7^Q=4\sqrt{2}$ (dashed blue online), respectively. All the plotted quantities are dimensionless.\label{figdisp}}
\end{figure*}

A gallery illustrating the Svetlichny parameter $S_n$ as a function of the phase space settings $p_0$ and $p_1$ is presented in Fig.~\ref{figeva} for some representative choices of  $n$, at a fixed value of the state parameter $a$. The plots show that for any $n$ (and sufficiently large $a$) there exist regions of phase space settings leading to a violation of the Svetlichny inequality (\ref{sveltineqn}) for the states under consideration. With increasing $n$, even more islands in the parameter space appear that enable such a violation. However, to further investigate the points of maximal violation and to comment on the dependence of the resulting $S_n^{\text{opt}}(a)$ on $n$ and $a$, as shown in Fig.~\ref{figdisp}, we need to distinguish between the cases of even and odd $n$.

For odd $n \geq 3$, motivated by the evident symmetry in the distribution of the peaks in Fig.~\ref{figeva}(top), one can verify that the antisymmetric setting $p_0 = \tilde{p}_n(a) = -p_1$ is always an admissible solution for Eqs.~(\ref{buqings}), which reduce to a single equation whose solution gives the optimal $\tilde{p}_n(a)$ (where the subscript denotes the number of modes, rather than the measurement setting). Numerics confirm that such a solution leads to the largest value of the Svetlichny parameter for odd $n$ in the considered states under displaced parity measurements, i.e., $S^{\text{opt}}_n (a) = S_n(a; 0,0,\tilde{p}_n(a),-\tilde{p}_n(a))$. Under these premises,  the resulting $S^{\text{opt}}_n(a)$ is plotted as a function of $a$ in Fig.~\ref{figdisp}(b). As clear from the inset of the Figure, one finds that there exists, for any odd $n$, a threshold value $\tilde{a}_n$ of $a$ such that $\tilde{p}_n(a)=0$ and $S^{\text{opt}}_n (a) = 1$ for $1 \leq a \leq \tilde{a}_n$, meaning that no genuine $n$-partite nonlocality can be detected below the threshold using displaced parity measurements, despite the fact that pure permutationally invariant Gaussian states are fully inseparable for any $n$ as soon as $a>1$ \cite{network,telepoppy,ghzw,contangle}. This was already noted in \cite{samy} in the case $n=3$. The threshold value $\tilde{a}_n$ to violate the Svetlichny inequality, as well as the optimal setting $\tilde{p}_n(a)$ to reach the largest violation provided $a>\tilde{a}_n$, can be determined analytically in principle by solving Eqs.~(\ref{buqings}), even though the problem becomes quite untractable for large $n$. For instance, for $n=3$ we get $\tilde{a}_3 = \sqrt{\frac32}$ and
\begin{equation}\label{popt3}
\tilde{p}_3(a)=\sqrt{\frac{1}{8z^+_3(a)}\ln \left[ \frac{a+2z^+_3(a)}{3a-2z^+_3(a)}\right]}\,,
\end{equation}
in agreement with the results of \cite{samy}\footnote{Note that there was a typo in the expression corresponding to  $\tilde{p}_3(a)$ in \cite{samy}, while Eq.~(\ref{popt3}) gives the correct formula.} However, and quite interestingly, a numerical evaluation reveals that $\tilde{a}_n$ quickly shrinks towards $1$ with increasing $n$ [see Fig.~\ref{figdisp}(b)], which suggests that almost all fully inseparable Gaussian states of the studied class, in case of a large odd number $n \gg 1$ of modes, exhibit a violation of local realism with the adopted measurements.

For even $n \geq 2$ (including the bipartite case $n=2$, when the Svetlichny parameter $S_2$ reduces to the Bell-CHSH one $M_2$), as apparent by the slight skewness in the islands of Fig.~\ref{figeva}(bottom), the setting $p_0=-p_1$ is not anymore a solution of Eqs.~(\ref{buqings}), which means that an optimization over two parameters remains to be performed, to obtain $S^{\text{opt}}_n (a) = \max_{\{p_{0,1}\}} S_n(a; 0,0,p_0,p_1)\,.$ Analytical expressions, if available, are quite cumbersome in this case, so one can comfortably resort to a numerical solution. The resulting $S_n^{\text{opt}}(a)$ is plotted as a function of $a$ in Fig.~\ref{figdisp}(a). As the inset of the Figure shows, and as numerical calculations confirm, in the case of even $n$ there is no threshold for the violation of the Svetlichny inequality, that is, $S_n^{\text{opt}}(a)>1$ for all $a>1$, revealing genuine $n$-partite nonlocality as soon as the Gaussian states under consideration are fully inseparable.

Finally, by comparing the two cases of even and odd $n$, i.e.~by a justaposition of the two panels of Fig.~\ref{figdisp}, we observe that $S^{\text{opt}}_{2k}(a) \geq S^{\text{opt}}_{2k+1}(a)$ for any $k \geq 1$, even though the difference between consecutive even and odd cases vanishes asymptotically for $a \gg 1$. Most importantly, however, for any $n$ the maximum Svetlichny parameter achievable with the considered measurements stays well below the maximum value $S_n^Q$ allowed by quantum mechanics, given by Eq.~(\ref{sveltcireln}) and indicated by dashed lines in Fig.~\ref{figdisp}. For instance, for $n=2,3$ we get $\lim_{a \rightarrow \infty} S^{\text{opt}}_n(a) = 4 \times 3^{-9/8} \approx 1.162$ \cite{jeong,samy}, while the maximum quantum violation amounts to $S_n^Q=\sqrt{2} \approx 1.414$. The conclusion we can draw from this extensive analysis is that one can feasibly detect genuine $n$-partite nonlocality by displaced parity measurements, but such observables are not sensitive enough to reveal an extremal violation of local realism in $n$-mode Gaussian states.

\section{Maximum tripartite nonlocality with pseudospin measurements}\label{sec:Pseudo}

In the original discussion by Einstein, Podolsky, and Rosen \cite{EPR35}, the idealized eigenstate of relative position and total momentum of two particles was argued to possess paradoxical nonlocal properties. In contemporary terms, we can say that such a continuous variable state (which is not normalizable, hence unphysical) is maximally entangled, i.e., it is characterized by a diverging entanglement entropy. Gaussian two-mode squeezed states, generated e.g.~by optical parametric amplifiers \cite{barnett}, approach such an ideal limit asymptotically in the regime of large squeezing. For this reason, it is natural to expect that the nonlocality exhibited by these states would reach the maximum allowed by quantum mechanics in the limit of infinite squeezing. This was in fact proven by showing that the bound (\ref{chshcirel}) for the CHSH parameter can be asymptotically saturated by such states, when using {\it pseudospin measurements} \cite{chen}.

In the following we show that, by means of optimized pseudospin measurements, the Svetlichny inequality can also be maximally violated on a class of pure permutationally invariant three-mode Gaussian states, up to the limit in (\ref{sveltcirel}).  This shows that continuous variable Gaussian states can display extremal genuine tripartite quantum nonlocality, which could not be revealed by using displaced parity operators. We emphasize that the measurements considered in this section require us to work directly in the Fock basis, thus losing some of the elegance and compactness of the phase space formalism adopted above. As a result, extending this  study beyond three modes appears challenging at present.

For a single mode, the pseudospin observable $\hat{Z}$ is defined as \cite{chen}
\begin{equation}
\hat Z(\boldsymbol{\xi})=\cos\theta\, \hat Z_z+\sin\theta(e^{-i\varphi}\hat Z_++e^{i\varphi}\hat Z_-),\label{pseudospino}\,,
\end{equation}
where $\boldsymbol{\xi}\equiv(\theta,\varphi)$ defines the measurement setting. In the Fock basis $\{\ket{n}\}$ of a single mode, the three operators appearing in Eq.~\eqref{pseudospino} are defined as
\begin{align}
\hat Z_z&=\sum_{m=0}^{\infty}\Big(\proj{2m+1}-\proj{2m}\Big),\\
\hat Z_+&=\sum_{m=0}^{\infty}\kebra{2m+1}{2m},\\
\hat Z_-&=\hat Z_+^\dagger.
\end{align}
A scheme to implement pseudospin measurements on a sequence of two-level atoms resonantly interacting with a cavity mode was described in \cite{chen}.

Moving on to a tripartite scenario, given a three-mode state $\rho$, we define the correlation function associated with the measurement of pseudospin operators $\hat{Z}^j(\boldsymbol{\xi}^j_{x_j})$ on each mode $j\in\{a,b,c\}$ with respective settings ${\boldsymbol{\xi}}^j_{x_j}$ (where $x_j \in \{0,1\}$ labels once more two possible settings per mode) as
\begin{equation}\label{pseudocorre}
\langle a_{x_a} b_{x_b} c_{x_c} \rangle = \tr \left[ \rho \, \hat{Z}^a(\boldsymbol{\xi}^a_{x_a}) \hat{Z}^b(\boldsymbol{\xi}^b_{x_b}) \hat{Z}^c(\boldsymbol{\xi}^c_{x_c}) \right]\,.
\end{equation}
Inserting the expression (\ref{pseudocorre}) into Eq.~(\ref{svelt}), we can then construct the Svetlichny parameter $S_3$ corresponding to pseudospin observables.

We consider a family of pure permutationally invariant three-mode Gaussian states with wavefunction
\begin{equation}
	\ket\psi=\frac{1}{\sqrt{\cosh r}}\exp\left({\frac{\tanh r}{2}\left(\frac{\hat a^\dagger+\hat b^\dagger+\hat c^\dagger}{\sqrt3}\right)^2}\right)\ket{000},\label{squizzone}
\end{equation}
where $\hat{j}$ is the annihilation operator on mode $j$, and $r>0$ plays the role of a three-mode squeezing parameter. These continuous variable GHZ-like states may be obtained by mixing a single-mode squeezed state \cite{barnett} with two vacuum states at a balanced ``tritter'' \cite{zukowski,network}. Their covariance matrix is local unitarily equivalent to the normal form given in Eq.~(\ref{sigma}) with an effective parameter $a \equiv \sqrt{\det{\boldsymbol{\alpha}}}=\frac{1}{3} \sqrt{5+4 \cosh (2r)}$.

\begin{figure}[t!]
\includegraphics[width=8cm]{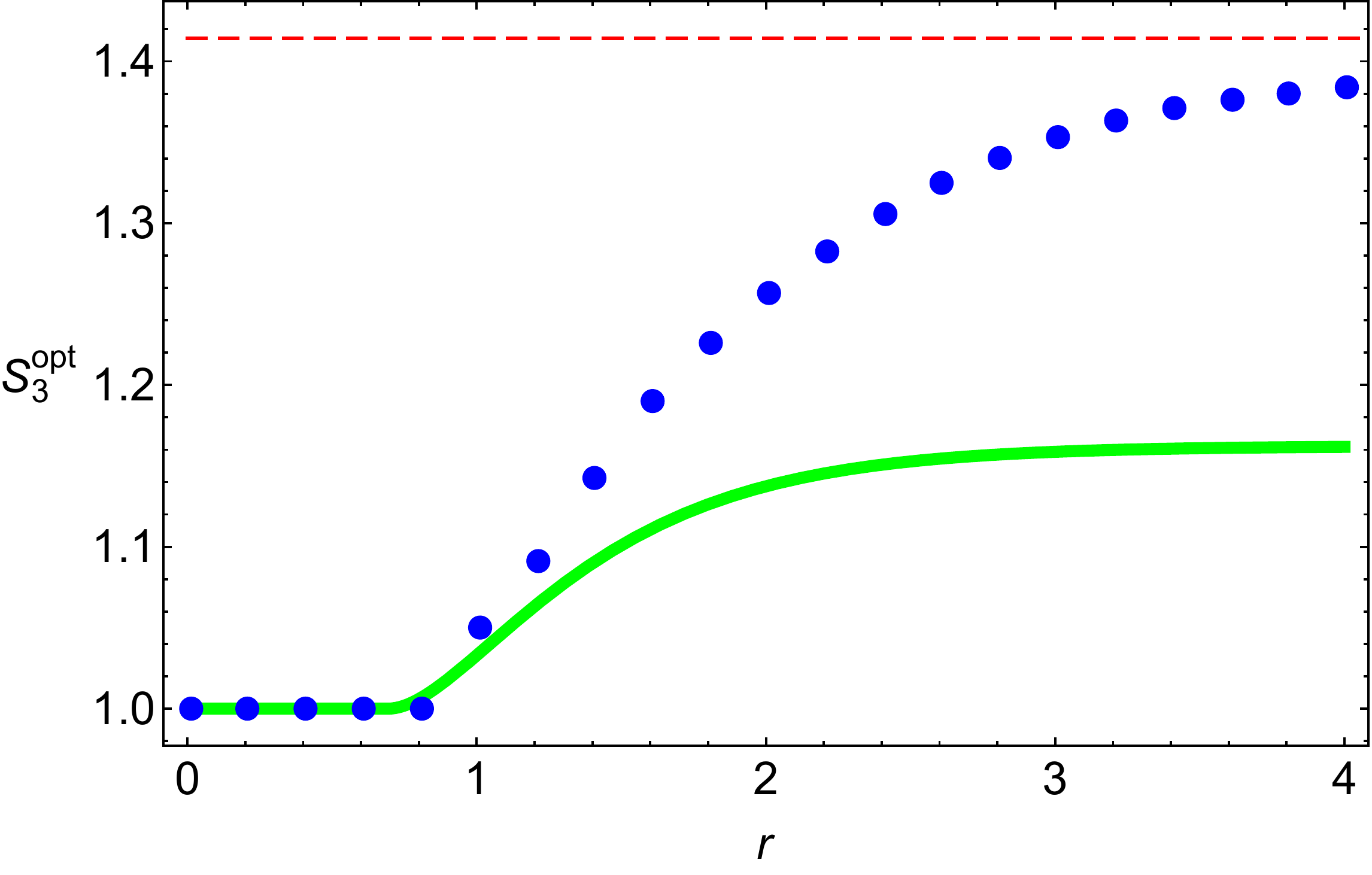}
\caption{(Color online) Optimal Svetlichny parameter $S_n^{\text{opt}}$ for the $3$-mode Gaussian states of Eq.~(\ref{squizzone}), using pseudospin observables (blue points) and displaced parity measurements (solid green line), plotted  versus the squeezing parameter $r$. As in Fig.~\ref{figdisp},  the dashed horizontal (red) line indicates the maximum value $S_3^Q=\sqrt 2$ of the Svetlichny parameter allowed by quantum mechanics. All the plotted quantities are dimensionless.\label{figatrita}}
\end{figure}

In Fig.~\ref{figatrita}, we plot the Svetlichny parameter $S_3^{\text{opt}}(r)$ for these states, optimized numerically over all the pseudospin measurement settings $\{{\boldsymbol{\xi}}^j_{x_j}\}$. Quite interestingly, we see that, in the limit of large $r$, the violation of the Svetlichny inequality in the considered scenario appears to approach the maximum quantum bound $S_3^Q=\sqrt{2}$. The value of the Svetlichny parameter for the same states using displaced parity measurements (optimized analytically as described in the previous section) is also plotted for comparison. As already remarked, the maximum achievable value of $S_3$ for three-mode Gaussian states using displaced parity measurements is  only $\approx 1.162$, and the states of Eq.~(\ref{squizzone}) reach this limit for $r \gg 1$ using such measurements.

We would like to go beyond the numerical analysis to provide a more rigorous evidence for the maximum tripartite quantum nonlocality of the Gaussian states under study. To this aim, we identify specific pseudospin measurement settings, given by
\begin{align}\label{settino}
	(\theta^a_0,\varphi^a_0,\theta^b_0,\varphi^b_0,\theta^c_0,\varphi^c_0)&=(0,\tfrac{\pi}{2},\tfrac{\pi}{4},\tfrac{\pi}{2},0,-\tfrac{\pi}{2}), \nonumber\\
& \\ (\theta^a_1,\varphi^a_1,\theta^b_1,\varphi^b_1,\theta^c_1,\varphi^c_1)&=(\tfrac{\pi}{2},\tfrac{\pi}{2},\tfrac{3\pi}{4},\tfrac{\pi}{2},-\tfrac{\pi}{2},-\tfrac{\pi}{2})\,. \nonumber
\end{align}
With this choice, and after straightforward yet somewhat tedious algebra, we obtain the following expression for the Svetlichny parameter:
\begin{align}
	S_3 & =\frac{\sqrt{2}}{4}+\frac{3\sqrt{2}}{2}{\rm Re}\Big(\langle\psi|\hat Z_{z}^{a}\hat Z_{+}^{b}\hat Z_{-}^{c}|\psi\rangle-\langle\psi|\hat Z_{z}^{a}\hat Z_{+}^{b}\hat Z_{+}^{c}|\psi\rangle\Big).\label{Sparticular}
\end{align}
Eq.~\eqref{Sparticular} can be derived by invoking the permutational invariance of the states  $\ket{\psi}$, and observing that any such state is an eigenstate of the total parity $\hat \Pi=-\hat Z_z^a\hat Z_z^b\hat Z_z^c$ with eigenvalue $+1$ (that is, it is a superposition of states with an even number of photons). Noting that, for any $j\in\{a,b,c\}$, $\hat Z_z^j$ preserves the total parity of a state, while $\hat Z^j_+$ and $\hat Z_-^j$ act respectively as raising and lowering operators for $\hat \Pi$, we obtain that the only nonzero correlation functions contributing to $S_3$ are $\bra{\psi}\hat Z_z^a\hat Z_z^b\hat Z_z^c\ket{\psi}=-\bra\psi\hat \Pi\ket{\psi}=-1,$ which does not depend on the squeezing parameter $r$, together with the squeezing-dependent quantities $\langle\psi|\hat Z_{z}^{a}\hat Z_{+}^{b}\hat Z_{-}^{c}|\psi\rangle$ and $\langle\psi|\hat Z_{z}^{a}\hat Z_{+}^{b}\hat Z_{+}^{c}|\psi\rangle$ and their complex conjugates. All the other nonzero correlation functions are obtained from these by permutation of the indices $(a,b,c)$. Furthermore, it is easy to spot from Eq.~\eqref{squizzone} that the expansion coefficients of $\ket{\psi}$ in the Fock basis are purely real, which implies that the correlation functions appearing in Eq.~\eqref{Sparticular} are real as well.

Our expression for $S_3$ can be further simplified noting that the total parity of $\ket\psi$ implies $\langle\psi|\hat Z_{z}^{a}\hat Z_{+}^{b}\hat Z_{-}^{c}|\psi\rangle =\langle\psi|\hat Z_{+}^{b}\hat Z_{-}^{c}|\psi\rangle$, while $\langle\psi|\hat Z_{z}^{a}\hat Z_{+}^{b}\hat Z_{+}^{c}|\psi\rangle=-\langle\psi|\hat Z_{+}^{b}\hat Z_{+}^{c}|\psi\rangle$. Then, Eq.~\eqref{Sparticular} may be cast in the elegant form
\begin{align}
S_3 & =\frac{\sqrt{2}}{4}\left(1+3\langle\psi|\hat Z_{x}^{b}\hat Z_{x}^{c}|\psi\rangle\right),\label{Simple}
\end{align}
where $\hat Z_x^j= \hat Z^j_+ +\hat Z^j_-$. It is thus clear that the state $\ket\psi$ would yield maximum violation of the Svetlichny inequality if it could satisfy the condition  $\langle\psi|\hat{Z}_{x}^{b}\hat{Z}_{x}^{c}|\psi\rangle=1$. This can only be achieved if $\ket{\psi}$ is a $+1$ eigenstate of $\hat{Z}_{x}^{b}\hat{Z}_{x}^{c}$. In the remainder of this section, we report extensive evidence that this is indeed the case in the infinite squeezing limit $r\to\infty$.
More precisely, we shall present a semi-analytical proof, supplemented by numerical evidence, that the following limit holds,
\begin{align}
\lim_{r\to\infty}\left\|\ket{\psi}-\hat Z_x^b\hat Z_x^c\ket\psi\right\|=0.\label{limit}
\end{align}

To begin with, we take a series expansion of the exponential in Eq.~\eqref{squizzone}, which yields
\begin{align}
\ket\psi=\frac{1}{\sqrt{\cosh r}}\sum_{n=0}^{\infty}\frac1{n!}\left(\frac{\tanh r}{6}\right)^n\left(\hat a^\dagger+\hat b^\dagger+\hat c^\dagger\right)^{2n}\ket{000}.
\end{align}
This allows us to find the expansion of $\ket \psi$ in the Fock basis,
\begin{align}
\ket\psi&=\frac{1}{\sqrt{\cosh r}}\sum_{n=0}^{\infty}\frac1{n!}\left(\frac{\tanh r}{6}\right)^n\nonumber\\
&\times\sum_{k_1+k_2+k_3=2n}\frac{(2n)!}{\sqrt{k_1!k_2!k_3!}}\ket{k_1,k_2,k_3}.
\end{align}

Our next step is to evaluate the Fock basis expansion of $\hat Z_x^b\hat Z_x^c\ket{\psi}$. This is easily done by recalling that the action of $\hat Z_x^b\hat Z_x^c$ on a Fock basis element is
\begin{equation}
	\hat{Z}_x^b \hat{Z}_x^c\ket{k_1,k_2,k_3}=\left\{\begin{array}{lr}
\ket{k_1,k_2+1,k_3+1} & \quad k_2 \text{ even, }k_3 \text{ even;} \\
\ket{k_1,k_2-1,k_3-1} & k_2 \text{ odd, }k_3 \text{ odd;}\\
\ket{k_1,k_2+1,k_3-1} & k_2 \text{ even, }k_3 \text{ odd;}\\
\ket{k_1,k_2-1,k_3+1} & k_2 \text{ odd, }k_3 \text{ even.}\\
	\end{array}\right.
\end{equation}

Fixing $k_1+k_2+k_3=2n$, we thus obtain the expansion coefficients
\begin{align}	&\bra{k_1,k_2,k_3}\hat{Z}_x^b\hat{Z}_x^c\ket{\psi}=\frac{1}{\sqrt{\cosh r}}\frac{(2n)!}{n!}\left(\frac{\tanh r}{6}\right)^n \nonumber\\
	&\times\left\{\begin{array}{lr}
	\left(\frac{\tanh r}{3}\right)\frac{2n+1}{\sqrt{k_1!(k_2+1)!(k_3+1)!}} & \quad k_2 \text{ even, }k_3 \text{ even;} \\
	\left(\frac{3}{\tanh r}\right)\frac{1}{(2n-1)\sqrt{k_1!(k_2-1)!(k_3-1)!}} & k_2 \text{ odd, }k_3 \text{ odd;}\\
	\frac{1}{\sqrt{k_1!(k_2+1)!(k_3-1)!}} & k_2 \text{ even, }k_3 \text{ odd;}\\
	\frac{1}{\sqrt{k_1!(k_2-1)!(k_3+1)!}} & k_2 \text{ odd, }k_3 \text{ even.}
	\end{array}\right.
	\end{align}

Thanks to the orthonormality of the Fock states, we can hence write
\begin{align}
	\left\|\ket{\psi}-\hat{Z}_x^b\hat{Z}_x^c\ket{\psi}\right\|^2=\frac{1}{\cosh r}\sum_{n=0}^\infty\sum_{k_1+k_2+k_3=2n}R_{k_1,k_2,k_3},
\end{align}
where
\begin{align}
&R_{k_1,k_2,k_3}=\left(\frac{(2n)!}{n!}\right)^2\left(\frac{\tanh r}{6}\right)^{2n}\\
&\!\!\times\left\{\!\begin{array}{lr}
\left(\frac{1}{{\sqrt{k_1!k_2!k_3!}}}-\left(\frac{\tanh r}{3}\right)\frac{2n+1}{\sqrt{k_1!(k_2+1)!(k_3+1)!}} \right)^2&\!\!   k_2 \text{ even, }k_3 \text{ even;} \\
\left(\frac{1}{{\sqrt{k_1!k_2!k_3!}}}-\left(\frac{3}{\tanh r}\right)\frac{1}{(2n-1)\sqrt{k_1!(k_2-1)!(k_3-1)!}}\right)^2 & k_2 \text{ odd, }k_3 \text{ odd;}\\
\left(\frac{1}{{\sqrt{k_1!k_2!k_3!}}}-\frac{1}{\sqrt{k_1!(k_2+1)!(k_3-1)!}}\right)^2 & k_2 \text{ even, }k_3 \text{ odd;}\\
\left(\frac{1}{{\sqrt{k_1!k_2!k_3!}}}-\frac{1}{\sqrt{k_1!(k_2-1)!(k_3+1)!}}\right)^2 & k_2 \text{ odd, }k_3 \text{ even.}\\
\end{array}\right. \nonumber
\end{align}

\begin{figure}[t!]
\includegraphics[width=8cm]{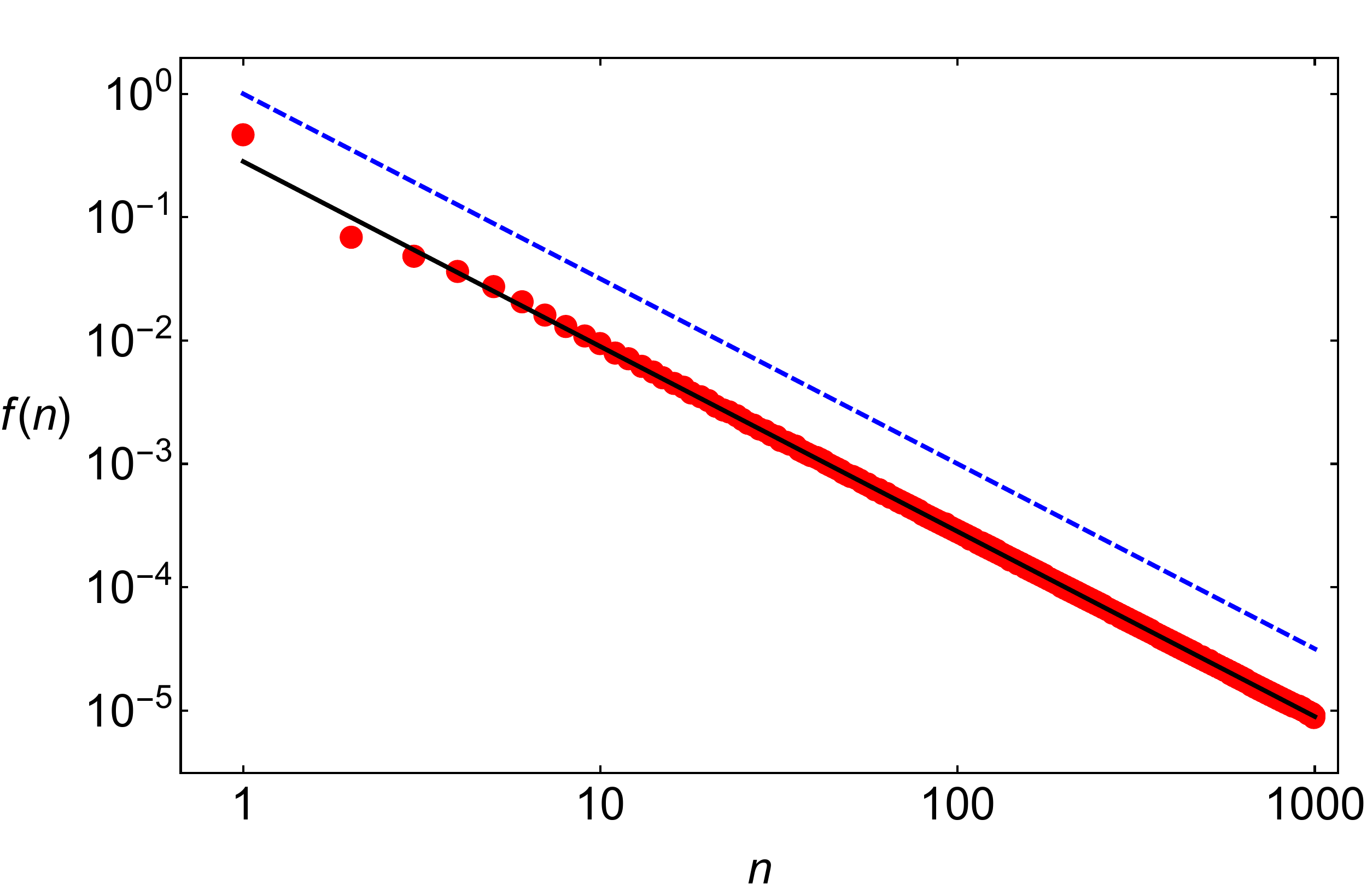}
\caption{(Color online) Numerical study of the sequence $f(n)$ as defined in Eq.~\eqref{sequoia}. Red dots indicate the numerically calculated value of $f(n)$. Notice how the asymptotic behaviour for large $n$ appears well approximated by $\simeq0.282\, n^{-3/2}$ (black solid line), which has been obtained via a power-law fit. The sequence $n^{-3/2}$ is also shown for comparison (blue dashed line). Logarithmic scale is used on both axes. All the plotted quantities are dimensionless. \label{numericazzo}}
\end{figure}

Since $\displaystyle\lim_{r\to\infty}\ [{\cosh(r)}]^{-1}=0$, in order for Eq.~\eqref{limit} to hold it would be sufficient to have $\sum_n\sum_{k_1+k_2+k_3=2n}R_{k_1,k_2,k_3}$ converging to a finite constant in the limit $r\to\infty$. This is for example the case if the sequence
\begin{equation}
	f(n)= \lim_{r\to\infty}\sum_{k_1+k_2+k_3=2n}R_{k_1,k_2,k_3},\label{sequoia}
\end{equation}
converges to zero faster than $n^{-(1+\epsilon)}$, for some $\epsilon>0$, in the limit $n\to\infty$. Our numerics confirm that this is indeed the case for $\epsilon=\frac12$, as shown in Fig.~\ref{numericazzo}. A numerical fit based on $n\lesssim1000$ yields the asymptotic behaviour $f(n)\simeq0.282\, n^{-3/2}$, which would provide a convergent sum.

Summing up, we have provided compelling evidence that the three-mode squeezed states in Eq.~\eqref{squizzone} asymptotically approach an eigenstate of $\hat Z_x^b\hat Z_x^c$ as $r\to\infty$. Correspondingly, this entails that in the same limit one would obtain a maximum quantum violation of the Svetlichny inequality using pseudospin operators with the settings of Eq.~(\ref{settino}), that is,
\begin{equation}
\lim_{r\to\infty}S_3(r)=\sqrt2.
\end{equation}

\section{Conclusions}\label{sec:Concl}

In this paper we have studied theoretically the degree of genuine multipartite nonlocality in pure permutationally invariant Gaussian states of $n$ bosonic modes, in terms of the largest amount by which the Svetlichny inequality \cite{svelt} is violated by specific measurements. When adopting displaced parity measurements \cite{kb}, we provided a prescription to find the optimal phase space settings in order to observe the most prominent violations of local realism, extending the results of \cite{samy}. These measurements nevertheless fail to reveal the maximum Svetlichny nonlocality allowed by quantum mechanics when operating on Gaussian states. For this reason we further considered pseudospin observables \cite{chen} and provided convincing evidence that such an ultimate bound is in fact attainable on Gaussian states when using these measurements, in particular in the three-mode instance. Extensions of this result to a higher number of modes, exploiting the symmetries of the states as outlined in our analysis, might be feasible, even though they appear significantly more intricate than the $n=3$ instance. Also in the case of pseudospin measurements, we identified particular settings which become optimal in the regime of large squeezing. These findings can be useful to guide an experimental demonstration of genuine multipartite continuous variable nonlocality for practical purposes.

Here we have focused on the violation of the Svetlichny inequality \cite{svelt}, which provides a sufficient condition for detecting genuine multipartite nonlocality. More recent studies have led to the identification of a larger set of weaker inequalities, whose violation (even without a violation of Svetlichny inequality) is still sufficient to demonstrate genuine multipartite nonlocality \cite{operationonloc,bancal}. Analysing these weaker yet more complex inequalities is considerably more cumbersome in continuous variable systems, and furthermore it is not clear a priori which inequalities can be violated on specific classes of states (and in some cases what is their  maximum possible quantum violation), even though the violation of one such inequality has been investigated theoretically for three-mode Gaussian states using displaced parity measurements in \cite{samy}. Here we were mainly concerned with identifying conditions to reveal the strongest possible signature of genuine multipartite quantum nonlocality, which justifies our focus on the Svetlichny inequality, and the pursuit of its maximum violation using Gaussian states. Whether such a quantitative violation might be interpreted operationally in terms of a figure of merit for  a continuous variable quantum information and communication task would be an interesting topic for further investigation.

In future work, it may also be worth extending our study to other correlations, such as quantum steering \cite{wiseman,wisemulti,armstrong}, a weaker and asymmetric form of nonlocality which can also be detected by the violation of suitable inequalities \cite{steeringreview1,steeringreview2}. Very recently, displaced parity and pseudospin observables have been considered to detect steerability of bipartite Gaussian states \cite{nogaa1,nogaa2}, and proven useful to reveal a larger set of steerable states than what can be characterized by using Gaussian measurements alone \cite{wiseman,kogiak}. Identifying the boundaries of the sets of steerable or nonlocal Gaussian states (in bipartite as well as multipartite continuous variable systems), and the maximum allowed violations of corresponding inequalities when acting with specific classes of feasible measurements, would be helpful to identify optimal resources for fully or partially device-independent quantum communication using continuous variable systems. We hope the present work can serve as a stimulus to further progress in such directions.

\begin{acknowledgments}
We warmly thank Antony R. Lee for his contributions to the earlier stages of this project. We acknowledge further discussions with Samanta Piano, Ioannis Kogias, and Ajit Iqbal Singh. This work was supported by the European Research Council (ERC) Starting Grant GQCOP (Grant No.~637352) and by the Foundational Questions Institute (fqxi.org) Physics of the Observer Programme (Grant No.~FQXi-RFP-1601).
\end{acknowledgments}

\end{document}